\documentclass{spie}

\usepackage{graphicx}
\usepackage[hidelinks]{hyperref}
\usepackage{float}
\usepackage{footnote}
\usepackage{listings}

\authorinfo{Send correspondence to I. Barclay, E-mail: BarclayIS@cardiff.ac.uk \\
Copyright 2019, Society of Photo‑Optical Instrumentation Engineers (SPIE). One print or electronic copy may be made for personal use only. Systematic reproduction and distribution, duplication of any material in this publication for a fee or for commercial purposes, and modification of the contents of the publication are prohibited.}

\title{A Conceptual Architecture for Contractual Data Sharing in a Decentralised Environment}
\author{Iain Barclay\supit{a}, Alun Preece\supit{a}, Ian Taylor\supit{a},\\
Dinesh Verma\supit{b}
\skiplinehalf
\supit{a}Crime and Security Research Institute, Cardiff University, Cardiff, UK\\
\supit{b}IBM TJ Watson Research Center, 1110 Kitchawan Road, Yorktown Heights, NY 10598, USA\\
}

\begin{document} 
\maketitle 

\begin{abstract}
Machine Learning systems rely on data for training, input and ongoing feedback and validation. Data in the field can come from varied sources, often anonymous or unknown to the ultimate users of the data. Whenever data is sourced and used, its consumers need assurance that the data accuracy is as described, that the data has been obtained legitimately, and they need to understand the terms under which the data is made available so that they can honour them. Similarly, suppliers of data require assurances that their data is being used legitimately by authorised parties, in accordance with their terms, and that usage is appropriately recompensed. Furthermore, both parties may want to agree on a specific set of quality of service (QoS) metrics, which can be used to negotiate service quality based on cost, and then receive affirmation that data is being supplied within those agreed QoS levels. Here we present a conceptual architecture which enables data sharing agreements to be encoded and computationally enforced, remuneration to be made when required, and a trusted audit trail to be produced for later analysis or reproduction of the environment. Our architecture uses blockchain-based distributed ledger technology, which can facilitate transactions in situations where parties do not have an established trust relationship or centralised command and control structures. We explore techniques to promote faith in the accuracy of the supplied data, and to let data users determine trade-offs between data quality and cost. Our system is exemplified through consideration of a case study using multiple data sources from different parties to monitor traffic levels in urban locations.
\end{abstract}


\section{INTRODUCTION}
\label{sec:intro}

Data from multiple and varied sources can be combined to create new collective intelligence systems\cite{saunders2017governing} or data ecosystems\cite{oliveira2019investigations} which are used to gain insight into business operations, train and run machine learning and other AI systems. Creating hybrid data-driven systems can also yield new assets, such as cleaned and harmonised data, labelled data sets, and trained machine learning models, such that customers and end-users have the opportunity to become data or asset suppliers to other parties. 

In this context, it becomes increasingly important to have reliable traceability on the origins of each data source, what its scope is, and what its terms of use are. This is important for proving ownership and the rights to use the source data, as well being able to identify quality or supply problems and alert users to problems or seek redress when things go awry. As intermediate data processing yields new reusable assets, it becomes more difficult to link these back to the original data sources, which may be necessary to have assurance of quality, availability and rights of use. Each original data source will be accompanied by a license, governing its terms of use, along with any commitments on quality or terms of service. When data is used to create new assets, the license terms of the original may still apply, but can very easily become `detached' from the data, and not get passed along. Further, users downstream who rely on the results of assets built from intermediate data may not be aware of either the licensing terms of the source data, or the quality of service commitments which accompany the data. This could be problematic in an open data setting, for example, where there may be no promise of service quality, or in commercially licensed data, where the data owner might be unaware that her data is being further re-used. 

In manufacturing industries it has become standard practice to track a product through its life-cycle from origin as raw materials, through component assembly to finished goods in a store, with the relationships and information flows between suppliers and customers recorded and tracked using supply chain management (SCM) processes\cite{lambert1998supply}. In agriculture and food (agri-food) production, traceability through the supply chain enables a trace back from a product in a supermarket to identify not only the source farm, but the batch of foodstuff used, as well as other products in which the same batch of foodstuff has been used.

By describing collective intelligence systems in terms of a supply chain, we are able to identify data sources and other contributions which form the source components or are the results of intermediate processes, and identify stakeholders in the overall system. As new assets are created through the system and used elsewhere, mappings can be extended to show the new users and provide a means to trace the newly-created data asset back to its original source components and suppliers, as well as track forward to find other users and consumers of the data. 

In this paper we present two case studies where hybrid data is used, and use these scenarios to induce requirements for an architecture which supports the creation and maintenance of systems with traceable data. For each use case we identify the desired goal of the application (i.e., its reason for being), and find the actors and data flows, enabling us to visualise a supply chain for the scenario. We use our findings to motivate the design of an architectural framework which enables us to record the origins of data sources, along with licenses or terms of use, and other artifacts and documentation associated with the data. As new assets are created from the source data, the provenance of the original data source remains attached to the new assets, providing a trail back to the source. Our framework can be used to provide transparency and traceability in data deployments. 

The remainder of this paper is organised as follows: Section~\ref{sec:related} draws upon research from manufacturing industries and agri-food production to discuss traceability in the supply chain and methods employed for tracing raw materials, and considers related work in the use of supplementary artifacts to augment understanding of data ecosystems. Section~\ref{sec:usecases} considers two case studies, which illustrate data sharing in machine learning model training and use; Section~\ref{sec:architecture} presents an architecture which we believe is capable of meeting data traceability requirements drawn from the case studies; Section~\ref{sec:conclusion} concludes the paper, and identifies areas for further work.

\section{RELATED WORK}
\label{sec:related}
In considering supply chains in the agri-food industry, Opara\cite{opara2003traceability} defines traceability as `the collection, documentation, maintenance, and application of information related to all processes in the supply chain in a manner that provides guarantee to the consumer and other stakeholders on the origin, location and life history of a product as well as assisting in crises management in the event of a safety and quality breach.' This definition applies well to the requirements for traceability in data systems. Opara\cite{opara2003traceability} further identifies six important elements of traceability which combine to constitute an integrated food supply chain traceability system: product traceability, process traceability, genetic traceability, inputs traceability, disease and pest traceability and measurement traceability. Whilst these are not all directly applicable to a data system, parallels can readily be drawn such that a system to provide traceability in a data ecosystem would need to provide traceability on: products, processes, inputs, errors or corrupt data, and measurements.

In research using sensors to provide traceability on animals in the agri-food supply chain, Kelepouris et al\cite{kelepouris2007rfid} provide  helpful terminology for the direction of analysis of information flow. \textit{Tracing} is the ability to work backwards from any point in the supply chain to find the origin of a product (i.e., `where-from' relationships)\cite{petroff1991framework}. \textit{Tracking} is the ability to work forwards, finding products made up of given constituents (i.e., `where-used' relationships)\cite{petroff1991framework}. An effective traceability solution should support both tracing and tracking; providing effectiveness in one direction does not necessary deliver effectiveness in the other\cite{kelepouris2007rfid}. 

Jansen-Vullers, van Dorp, and Beulens\cite{jansen2003managing} and van Dorp\cite{van2003traceability} discuss the composition of products in terms of a Bill of Materials (BoM) and a Bill of Lots (BoL).
The BoM is the list of types of component needed to make a finished product of a certain type, whereas the BoL lists the actual components used to create an instance of the product. In other words, the BoM might specify 6 $\times$ M6 bolts, whereas the BoL would identify which batch the actual bolts used in the building of a particular assembly were part of. The BoM and BoL can be multi-level, wherein components can be used to create sub-assemblies which are subsequently used in several different product types. 

Data ecosystems typically span organisations, with one organisation taking the role of a keystone partner\cite{den2011innovation} and other stakeholders contributing data or other resources. Distributed ledgers, such as those afforded by blockchain technologies, provide a means of recording information and transactions between parties, such as organisations in an ecosystem, who do not have a formal trust relationship\cite{tapscott2017blockchain}. The design of a blockchain system ensures that data written cannot be changed, providing a level of immutability and non-repudiation which is well suited to keeping an auditable record of events and transactions which occur between organisations. The use of blockchain technologies to enhance traditional SCM environments is an emerging area of development and research\cite{kim2018toward}. Sermpinis and Sermpinis\cite{sermpinis2018traceability} discuss the benefits of implementing a supply chain traceability system on blockchain technology. They contend that the transparency and tamper-proof characteristics of the Bitcoin\cite{nakamoto2008bitcoin} blockchain implementation make it a suitable platform for storing data relating to a manufacturing supply chain in a decentralised way, such that no central party assumes ownership and has the ability to change the data records.

State-of-the-art blockchain platforms, such as the Ethereum Project\cite{wood2014ethereum}, allow for the deployment of so-called smart contracts, implementing Szabo's vision\cite{szabo1994smart} of a "computerised transaction protocol that executes the terms of a contract", and can be considered as "‘autonomous agents’ stored in the blockchain, encoded as part of a ‘creation’ transaction that introduces a contract to the blockchain"\cite{luu2016making}, thus allowing blockchain platforms to provide non-repudible dynamic behaviours alongside their immutable storage capabilities.

Much research has been conducted into data provenance, and the provision of metadata to describe the lineage of data, notably W3 PROV\cite{missier2013w3c}. Groth's\cite{groth2013transparency} proposal to record the provenance of data and digital information in a `fair trade certificate for data' is an example of the type of supporting information which brings valuable context for meaningful traceability in data ecosystems. Other assets which can be considered to be useful supplementary information, include Hind's proposed Supplier's Declaration of Conformity\cite{hind2018increasing} which provides an overview of an AI system, detailing the purpose, performance, safety, security, and provenance characteristics of the overall system. At the component level, Gebru et al\cite{gebru2018datasheets} explore the benefits of developing and maintaining `Datasheets for Data', which replicate the specification documents that often accompany physical components, and Mitchell et al\cite{mitchell2019model} propose a document format for AI model specifications and benchmarks. 

The US Department of Commerce are currently working on the NTIA Software Component Transparency initiative to provide a standardised Software BoM\footnote{\url{https://www.ntia.doc.gov/SoftwareTransparency}} format to detail the sub-components in software applications. The intent is to give visibility on the underlying components used in software applications and processes such that vulnerable out-of-date modules can easily be identified and replaced. Tools such as CycloneDX\footnote{\url{https://cyclonedx.org}}, SPDX\footnote{\url{https://spdx.org"}}, and SWID\footnote{\url{https://www.iso.org/standard/65666.html}} have defined formats for identifying and tracking such sub-components.

\section{CASE STUDIES}
\label{sec:usecases}

The supply chain model described by Lambert\cite{lambert1998supply} is used as a framework with which to describe the chosen use cases. The components of the supply chain are considered from the viewpoint of the focus organisation as having a first level supplier, a `supplier's supplier', or second-level supplier, a customer and potentially a second level customer or `customer's customer'. Supplier and customer can include internal and external sources.
Each case study is presented with an overview of its application area and its goal. The input and output data sources are identified, and mapped onto a supply chain. Further, the data sources and processes which create both intermediate assets and the final outcome of the systems are considered, and described in terms of a BoM.

\subsection{London Traffic Congestion Classification System}

The London Traffic Congestion Classification System (LTC-CS)\cite{harborne2018reasoning} has been designed to demonstrate how different networks and data sources can be combined to develop situational understanding in coalition environments. The system comprises a trained machine learning (ML) model and a reasoning system for categorising levels of traffic congestion in an urban environment. The demonstration makes use of image data and video data provided under an Open Data license by Transport for London (TfL), along with mapping data from OpenStreetMap (OSM), again provided under an Open Data license.

To provide a dataset to train the congestion classification model, static images of traffic scenes were retrieved via TfL URLs and labelled by human annotators, identifying whether each scene shows situations where traffic is considered to be congested or not congested. In order to increase the size of the training data set, some source images from the archive are duplicated and then manipulated by being cropped or flipped. 

In a second part of the system, video streams are scanned for moving objects, or blobs. Key frames of the video are then used to identify and determine the positions of cars, and the results are merged to find the previously detected moving blobs that occupy the same positions as the detected cars. The velocity of the objects identified as cars are compared to speed limits of the road section, which is taken from the OSM map data from the location of the video clips. The difference between the speed of the travelling traffic and the speed limit of the road is used to determine whether the road is congested or not. This part of the system is used as second source at inference time, to provide an additional viewpoint on the road's state from that provided by the classifier using the trained ML model.


\subsubsection{Data Sources and Outputs}
This use case is typical of a supervised machine learning environment, where the data used for training is static (archived), whereas the data used for runtime congestion classification is live (dynamic). The source data for training the machine learning model, along with the data ownership and licensing constraints is identified in the upper part of Table \ref{table:tfl_ipdata}, while the data used for the runtime part of the system is below.

\begin{savenotes}
\begin{table}[th]
\begin{center}
\begin{small}
\begin{tabular}{|l|l|l|l|}
\hline
\textbf{Source Data} & \textbf{Description}           & \textbf{Supplier} & \textbf{License}     \\ \hline
\multicolumn{4}{|c|}{\textbf{Training data}} \\ \hline
Traffic Images               & Road scene images & 
TfL               & Open Data\footnote{\url{https://tfl.gov.uk/info-for/open-data-users/}} \\ \hline
Image labels          & Classifications of images & Internal               & None\\ \hline
\multicolumn{4}{|c|}{\textbf{Runtime data}} \\ \hline
Traffic Images               & Road scene images & 
TfL               & Open Data (see above) \\ \hline
Video Clips          & Road scenes & TfL               & Open Data (see above)\\ \hline
Map data             & Speed limits for roads & OpenStreetMap   & Open Data\footnote{\url{https://www.openstreetmap.org/copyright}} \\ \hline
\end{tabular}
\end{small}
\caption{Data sources used by the LTC-CS}
\label{table:tfl_ipdata}
\end{center}
\end{table}
\end{savenotes}

\begin{figure}[th]
\centering
\includegraphics[width=0.9\textwidth]{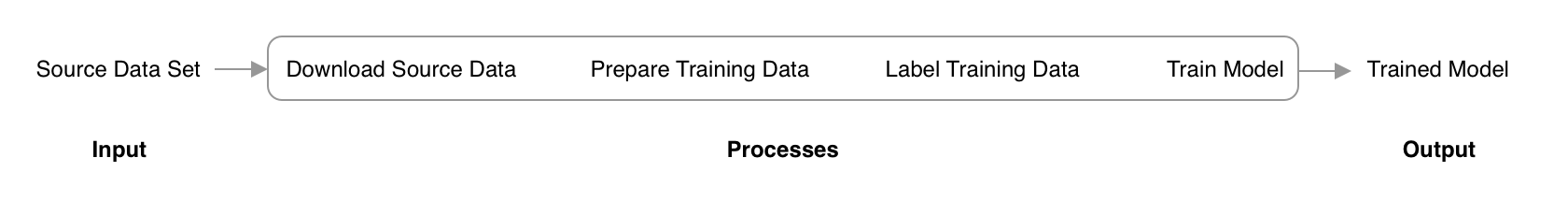} 
\caption{Work processes in LTC-CS Training}
\label{fig:tflprocesses}
\end{figure}

The image source data has had work applied to it to convert it into a suitable data set for training the machine learning model. This process involved cropping the source images to remove logos, and then creating a larger data set by duplicating and modifying images, e.g., by flipping copies of images. This increased the total volume of images suitable for use in training. Further human effort was applied to categorise and label the modified source images and to train the model, as illustrated in Figure~\ref{fig:tflprocesses}.

\begin{figure}[th]
\centering
\includegraphics[width=0.8\textwidth]{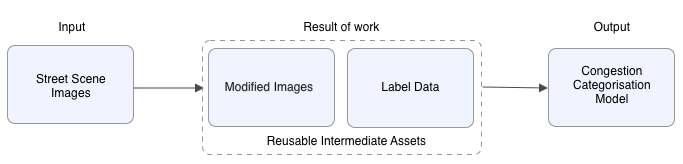} 
\caption{Assets produced by processing in TfL Scenario}
\label{fig:tflflow}
\end{figure}

As well as the trained model, the work processes create new assets, namely a modified version of the input data set and a set of labels for the this modified data set (Figure \ref{fig:tflflow}). Both these intermediate assets are potentially reusable in other scenarios, and provide good examples of a data production process producing new resources which could be used elsewhere in an organisation, shared with partners or even made commercially available. It is important to note, however, that the license of the original source data could have an impact on how these new assets are able to be reused.

The output assets arising from the work on the input data whilst training the model can be identified as:
\begin{description}
\item [A modified data set] The source data is modified in order to make it suitable for use in the training process. This process includes cleaning the source images by removing confusing elements, as well as duplicating and flipping some of the images to provide a larger training set.
\item [Labelled scenes] A second output from the model training process is the production of a set of labelled images, which has the potential for re-use in training other machine learning models. 
\item [A re-usable model] The algorithm and trained model for identifying traffic congestion from images of road scenes, could be reused in similar scenarios.
\end{description}

\subsubsection{The Data Supply Chain}
\begin{figure}[th]
\centering
\includegraphics[width=0.8\textwidth]{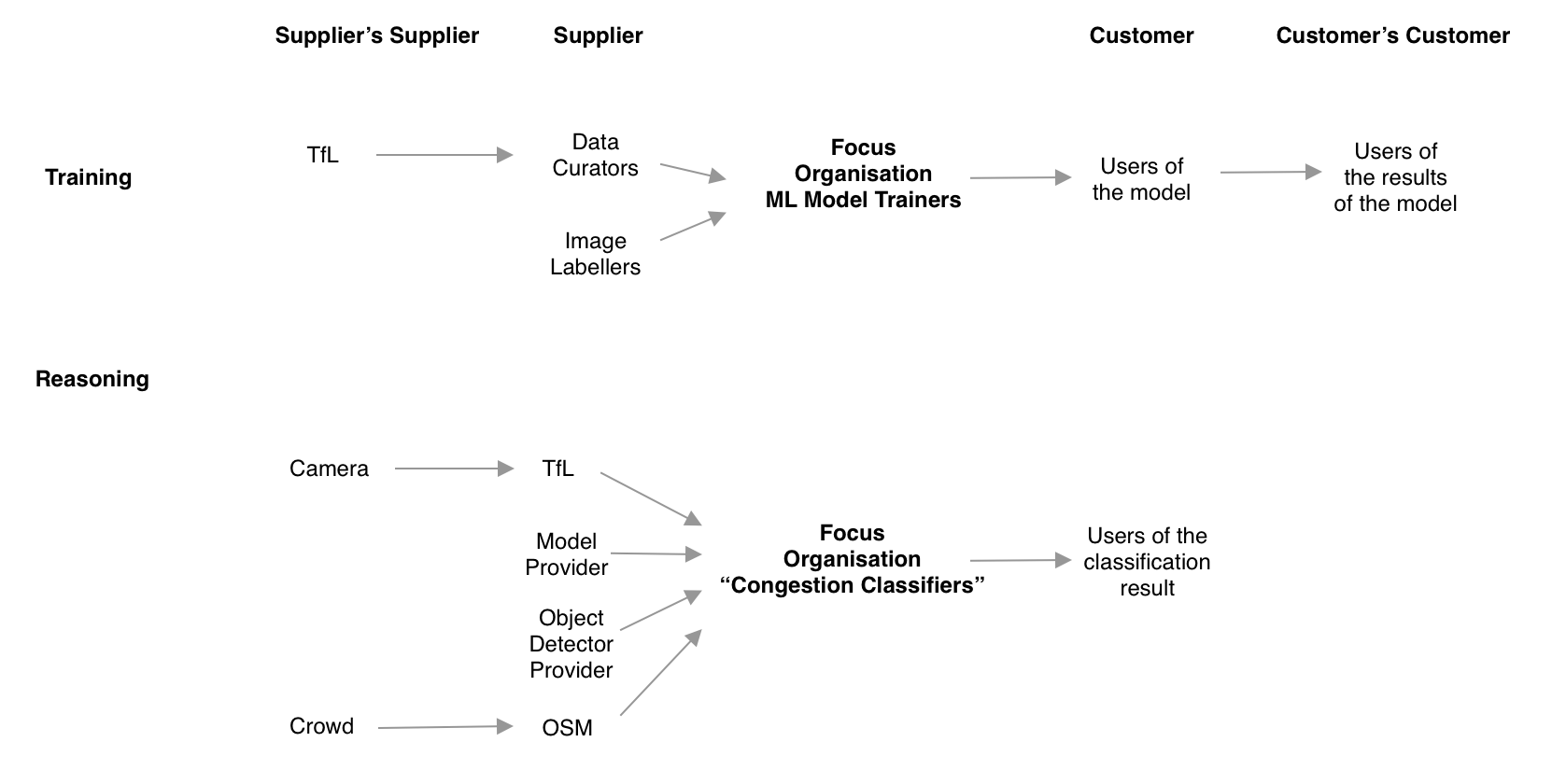}
\caption{LTC-CS Data Supply Chain}
\label{fig:tflsc}
\end{figure}

The LTC-CS case study uses data in two distinct processes - training of the model, and then inference using the model. The supply chain can be split into these two phases and illustrated as such (Figure \ref{fig:tflsc}), with the congestion predicting model produced as result of the training phase becoming an internally supplied input component in the inference phase.

The data supply chain shows a conceptual model for the system; it is important to note that any particular instantiation of the system would have data coming from different cameras (via different API requests) and different sections of the street map - for example, a request for congestion information on Oxford Street would use a different camera and section of OSM data from a request on Hyde Park Corner. Depending on the implementation, these different cameras could be accessed via different API URLs, or by passing a unique identifier into a single API URL. Similarly, the model and `object detector' code modules may be the same for distinct camera requests, or instances of the system could be deployed for each location supported. In either case, the version numbers of the model and object detector would change thorough the life of the system, and would be need to be recorded in a BoL for a particular decision. 

\begin{figure}[ht]
\centering
\includegraphics[width=0.8\textwidth]{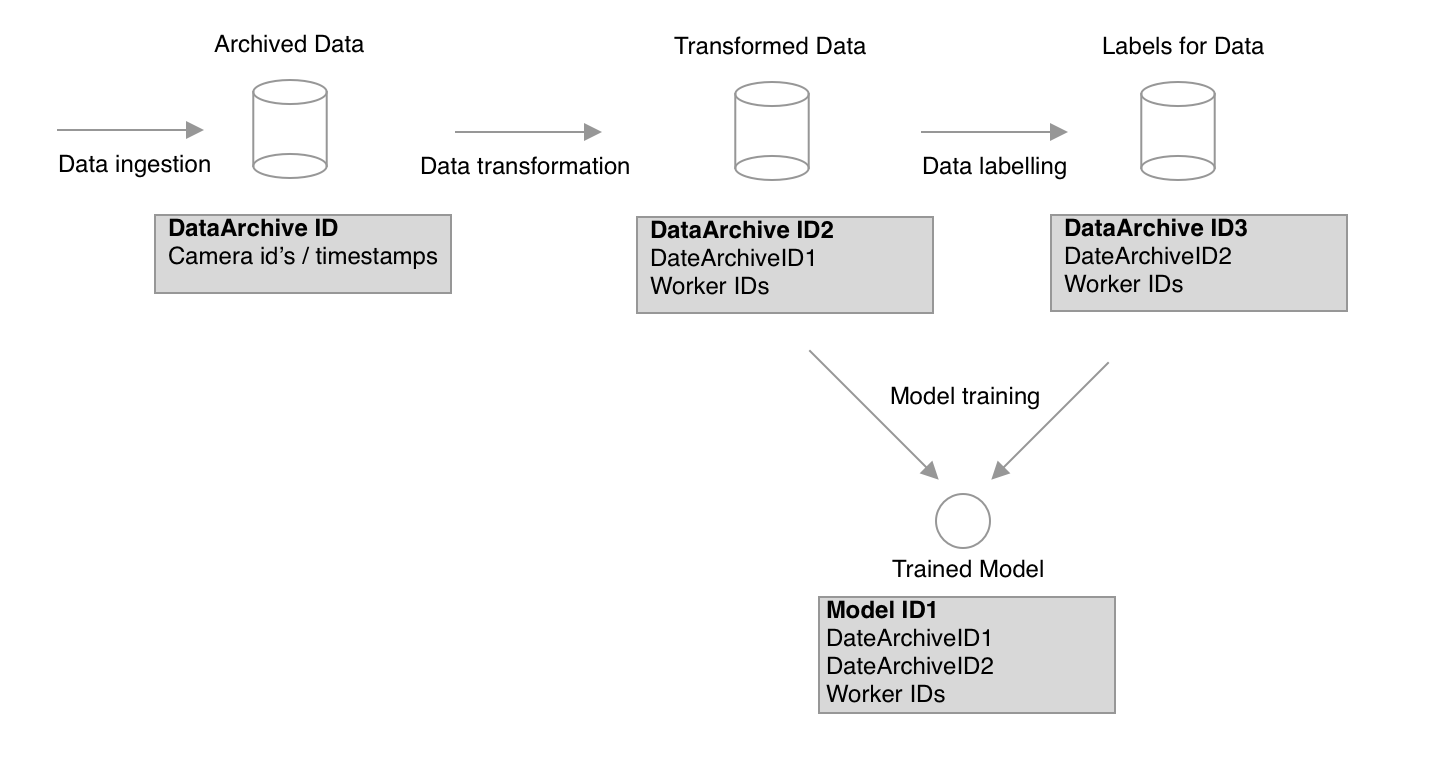} 
\caption{BoL for model training in the LTC-CS}
\label{fig:tfl_bol_model}
\end{figure}

\begin{figure}[ht]
\centering
\includegraphics[width=0.95\textwidth]{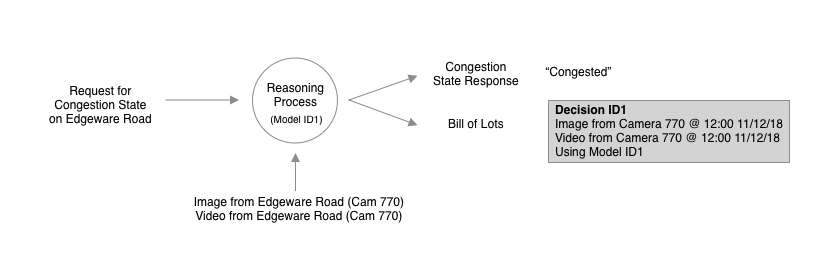} 
\caption{BoL for a runtime request in the LTC-CS}
\label{fig:tfl_bol1}
\end{figure}

Figure \ref{fig:tfl_bol_model} shows an example BoL for model training, and Figure \ref{fig:tfl_bol1} shows a BoL for a congestion status request, detailing the data and model used. Each instance of the model would have a unique identifier, and by tracing back its own BoL would be identified, such that the data and processes used to create it could be found. Similarly, from the model viewpoint, the BoL for any decision or other assets created from the model could be tracked.

Breaking out the data sources, models and other components employed in the inference process into constituents which describe a BoM for the system, and then generating a BoL for each instance or run which can accompany the output, makes it is possible to trace the data used in the reasoning process back to the source data used to train the model, and further to identify any ancestor data or workers involved in preparing the data. 

\subsection{Fusion AI}
During coalition operations, Intelligence, Surveillance and Reconnaissance (ISR) operations generate a substantial amount of data in a variety of modalities, including images, videos, acoustics, electromagnetic signatures that characterise different objects, entities and people that appear in the area of coalition operations. This data can be used to train AI models for a variety of tasks, such as differentiating insurgents from civilians, or  detecting enemy planes or assets from a distance.  

A common practice in coalition operations is for different partners to operate in different geographic areas, e.g. for a peace-keeping operation, the U.S. and UK may locate their forces in adjacent but different regions of a disturbed area. As a result, the training data collected by each member of the coalition are different, and combining them could lead to a significant improvement in the AI models that are created. However, sharing the data in the raw may not be always feasible among coalition partners. 

\begin{figure}[ht]
\centering
\includegraphics[width=0.5\textwidth]{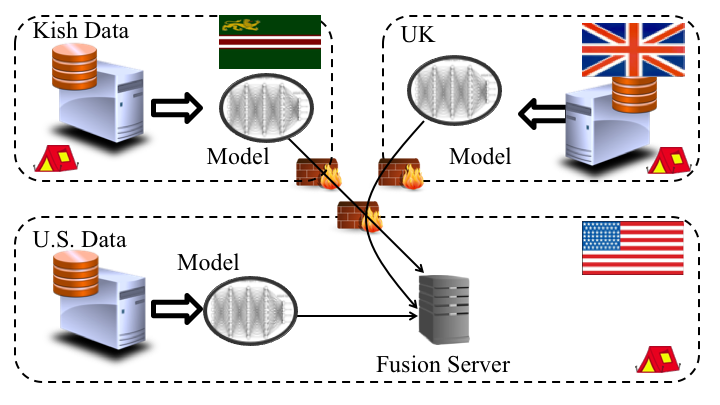} 
\caption{Scenario for Fusion AI for a coalition of three countries: U.S., UK and a hypothetical nation called Kish}
\label{fig:fusion-ai}
\end{figure}

Some of the restrictions on sharing the raw data come from limitations on network bandwidth that are available in tactical operations; for example, exchanging large amounts of data may not be viable over low-speed high-latency satellite links. 
Other type of restrictions preventing sharing of data come from sensitivity issues around the data that is collected. If a coalition partner has collected very high-fidelity images using some cutting edge technology, it may not want to reveal information about its capabilities to some other coalition partners. Similarly, sharing of some data may reveal sensitive information, such as the location of a camera or other sensor.

As such, coalition partners may benefit from federated learning. In this case, the partners train AI models on their local data, and share models with each other. The shared models can be fused or combined together to create a common model that captures the patterns from all of the training data. Different flavors of algorithms~\cite{juliervermacirin2018} for fusion can be constructed based on the type of trust among coalition partners, the level of coordination that can be permitted, and how the data is distributed across different coalition partners.

The goal of the Fusion AI system is to produce a shared AI model that can be used by all the entities that are involved in federated learning. The specific type of AI model would depend on the task being performed. In a coalition setting, a typical AI model may be one that recognises different types of people from images, or identifies potential insurgents which are differentiated from civilian population; or it may be a model to identify vehicles based on their acoustic or electromagnetic footprint. 

The data sources used in federated learning are similar, but they are provided by each of the individual entities participating in the federated learning process. The data source could be image data, or ISR data in the case of coalition operations. 
The output of the system is a shared machine learning model, which is built collaboratively across different sites. 

\begin{figure}[ht]
\centering
\includegraphics[width=0.47\textwidth]{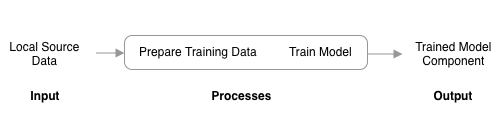}~~~~~~~~\includegraphics[width=0.41\textwidth]{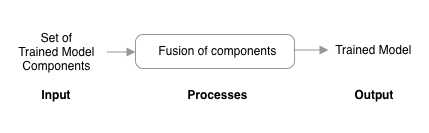}
\caption{Work processes for a single coalition member (left); Component models from coalition partners contribute to final model (right)}
\label{fig:fusionprocesses1}
\end{figure}

Considering the system workflow, each coalition partner will be responsible for gathering and preparing source data, and training their component of the complete model, as illustrated in Figure~\ref{fig:fusionprocesses1} (left).
This process will be replicated across contributing members of the coalition, with the set of trained model components from all partners being fused together in a final operation to create a trained model, which will be available for use in reasoning (Figure~\ref{fig:fusionprocesses1}, right). The fusion operation is not a fixed process, and introduces new elements which would need to be tracked in a supply chain model. Fusion may use rules or policies to determine how much of the model provided by each partner will be used, or what weighting to give it in the final fused model. Such rules would need to be captured, as they are significant factors in the provenance of the final fused model, and could change from time-to-time. There are a number of possible algorithms that can be used to create the fused model, and indeed, even the ordering of the component models as they are applied in the fusing process can make a difference to the outcome of the fused model. The algorithm used and any weighting factors or sequencing should be safely recorded to create the model's provenance.  

For each partner in the coalition, the work processes to create data suitable for training their element of the final model could lead to intermediate assets with additional value outside of the complete system. As seen in the traffic congestion scenario, such assets might include cleaned or processed variants of the source data, which could be re-used internally or sold or traded in future operations. Further, the intermediate elements of the models trained by each coalition partner as their contribution to the overall model could have some ongoing utility and value.

%

\begin{figure}[ht]
\centering
\includegraphics[width=0.8\textwidth]{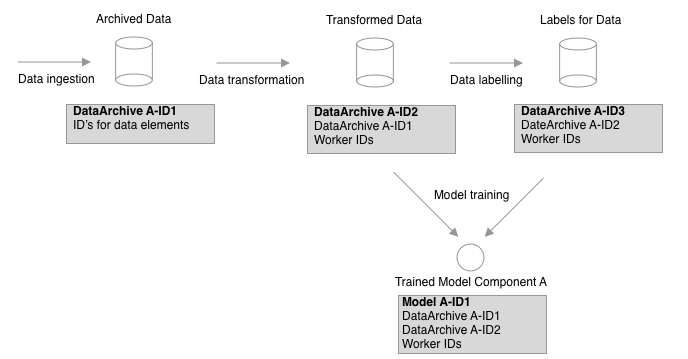}
\caption{BoL for coalition member $A$.}
\label{fig:coalitionA}
\end{figure}

\begin{figure}[ht]
\centering
\includegraphics[width=0.8\textwidth]{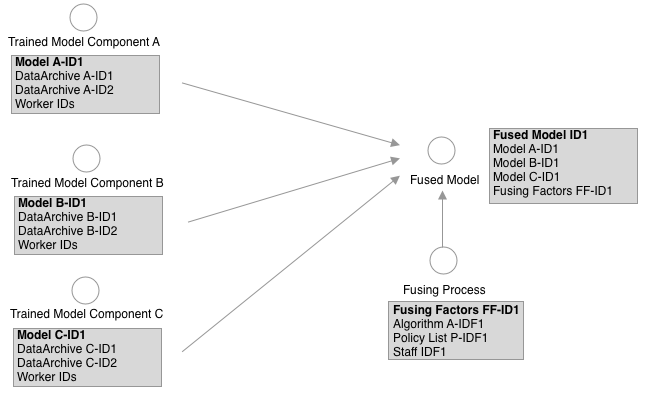}
\caption{BoL for final fused model, detailing contributions.}
\label{fig:fusionBOL}
\end{figure}

To produce a traceable BoL for a Fusion AI system, we need first to consider the BoM and BoL produced for the individual coalition contributors. Extending the processes of data ingestion, transformation and labelling as used in the traffic congestion case into each coalition member of a federated AI system, gives a per-coalition member BoM, which would create a BoL as illustrated in Figure~\ref{fig:coalitionA} for coalition member $A$. A similar intermediate BoM would be instantiated through coalition members $B$ and $C$, with different and unique identifiers used for each of the elements. As the model components are fused to create the overall model, a complete BoL can be generated, as illustrated in Figure~\ref{fig:fusionBOL}. As discussed, this aggregated BoL details the component models, any rules or policies employed in model selection and details of the fusing process and algorithm used (fusing factors), and allows for tracing back through the hierarchy to identify contributing components. When tracing back, it is possible to identify the component models, their source data, and the contributing staff, as appropriate. When the fused model is used in reasoning, its BoL is available, such that the fusing factors can be traced along with the contributing models and the data elements that seeded those models, as was shown in Figure \ref{fig:tfl_bol1}.

\section{CONCEPTUAL ARCHITECTURE}
\label{sec:architecture}

In this section, we present an architecture suitable for providing the levels of tracking and tracing needed in decentralised data ecosystems, such that data assets and uses of those assets can both be traced back to the origin and tracked forward to identify other users. The proposed solution defines a model  based on the BoM scheme, which is instantiated into a new BoL each time the deployment is run.

\subsection{Data Model}

The BoM data model describes the data supply chain in terms of being a collection of \textit{assemblies} 
each of which will typically have at least one data input, and can produce new data as its output. Data output from one assembly can be used as a data input in another assembly within the current BoM, or used in other systems, being referenced in their BoM. To reflect this, we refer to both data inputs and outputs as \textit{data sources}. Figures~\ref{fig:tflprocesses} and ~\ref{fig:tflflow} above identified processes which would be suitable for modelling as assemblies.

An assembly can also contain \textit{artifacts}, which is the term applied to any pertinent software components, ML models, and documentation such as licenses, policy documentation, Declarations of Conformity\cite{hind2018increasing}, datasheets for data\cite{gebru2018datasheets} or model specifications\cite{mitchell2019model}. By attaching artifacts to assemblies in the BoM definition we can ensure that each BoL retains a full record of its heritage and dependencies.

\begin{figure}[ht]
\centering
\includegraphics[width=0.75\textwidth]{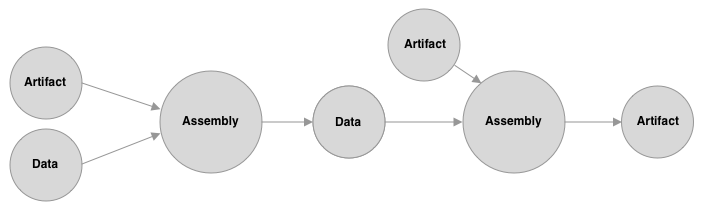}
\caption{Assemblies can be chained in a BOM}
\label{fig:simpleBOM}
\end{figure}

An assembly can produce a new artifact as its output; for example, an assembly which trained an AI model would produce the trained model as its output. The trained model would be considered an artifact which could be used in other assemblies. Figure~\ref{fig:simpleBOM} illustrates two such `chained' assemblies.

The BoM illustrated in Figure~\ref{fig:simpleBOM} could describe a simple AI model training process containing two assemblies. The first assembly representing the data labelling process, and the second the model training process. The BoM describes the system's data source and artifacts in abstract terms, and the BoL populates it for each run through of the process. The first input data source is the source training data, the artifact might be a roster of the staff employed to label the data, and the central data source (which, as illustrated, is both the output of the data labelling assembly and the input to the model training assembly) could be a labeled data set. The artifact in the second assembly would be relevant to the model training process, for example the parameters used in training. The output artifact would be a trained model.

\begin{figure}[ht]
\centering
\includegraphics[width=0.75\textwidth]{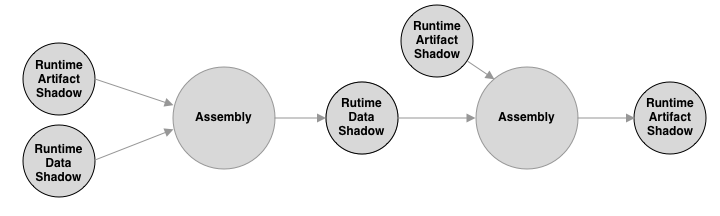}
\caption{BoLs contain runtime shadows for Data Sources and Artifacts.}
\label{fig:simpleBOL}
\end{figure}

The BoM defines a map of the structure of the system by providing a record of the connections between the assemblies, as well as any static data that applies to the contained data sources or artifacts. This static information could include a location for access to the data (e.g., an API URL, or address of a smart contract on a blockchain), acceptable data threshold levels or response requirements for active quality of service (QoS) monitoring. Each time the process described by the BoM is run, the application code should instantiate a new BoL for its BoM, which is used to store a record of the dynamic elements of each run. Figure~\ref{fig:simpleBOL} shows the structure of a BoL, and illustrates the shadow data that is used to record the dynamic state of the components in the instantiated BoM. 

\subsection{Simple Traffic Congestion Example}
\begin{figure}[ht]
\centering
\includegraphics[width=0.5\textwidth]{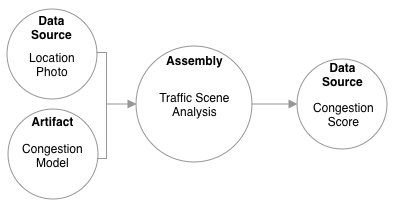}
\caption{The components of our example HPC-CS traffic congestion system}
\label{fig:HPC}
\end{figure}

By way of illustration, consider a simple software application --- which we'll call the Hyde Park Corner Congestion System (HPC-CS) --- that is a simplification of the previously described LTC-CS, providing a `congestion score' for a single fixed location, i.e., Hyde Park Corner, depending on how much traffic the application determines is currently at the location. This simple process has a single assembly, \textit{Traffic Scene Analysis}, an input data source \textit{Location Photo}, an ML model artifact \textit{Congestion Model} and an output data source \textit{Congestion Score} (Figure \ref{fig:HPC}).

In defining the BoM for the HPC-CS congestion scoring process, each element can be given a name and description, and static elements can be declared; for example, the URL to be used to retrieve a live photo from the location of interest could be specified. 
As the application code runs, it should refer to its BoM (via its instantiated BoL) to get locations for data it needs to access, and write any required dynamic information to its BoL for permanent storage.

In the congestion monitoring example, the data source for the traffic scene holds a static URL for a live camera. The application code should retrieve this information, access the photo, and (if appropriate) store a permanent copy of the photo to its own archives. A reference to the location of the archived copy of the photo used in the run should be written to the shadow data item, and saved as part of the archival of the BoL, along with the resultant congestion score, which should be written to the shadow of the result data source item.
 
Thus, each data source and artifact in each BoL would have any dynamic values recorded and stored in a database as a persistent record of the run, so that each of the assemblies in the BoL would have traceable input and output data values which could be accessed in the future if needed.

\subsection{Decentralised Data Sharing}

The simple HPC-CS traffic congestion example described above can be expanded to illustrate how the BoM model could be used to support a decentralised data sharing engagement with several suppliers. Further, we can use a blockchain or distributed ledger to provide an immutable record of a request for data services from suppliers, as well as enforce quality of service requirements or provide direct payment to data providers as a data request is made.

\begin{figure}[ht]
\centering
\includegraphics[width=0.8\textwidth]{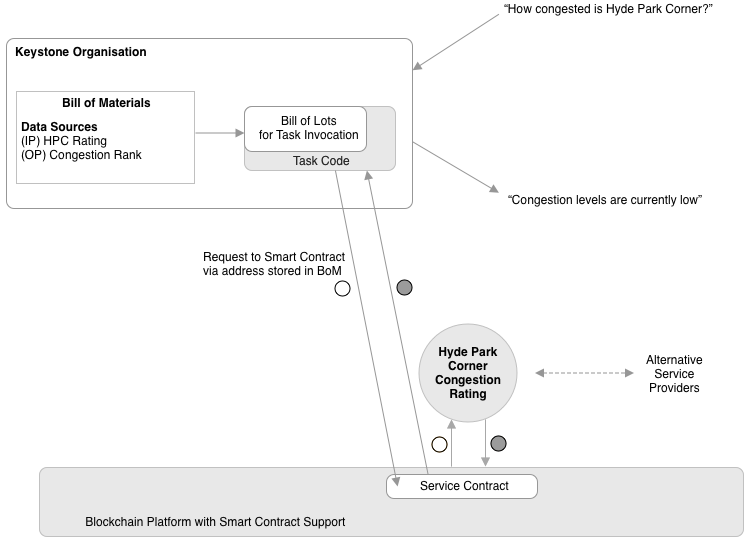}
\caption{The BoM model in a decentralised environment.}
\label{fig:decentralisedBOM}
\end{figure}

In the expanded scenario, the previously-described HPC-CS becomes a provider of a congestion rating to a larger system which ingests and processes the ratings from several data providers, as illustrated in Figure~\ref{fig:decentralisedBOM}.

The means by which the HPC-CS congestion scoring service is invoked is through a smart contract which is stored and runs on a blockchain platform, such as the public Ethereum blockchain. The address of the smart contract and metadata (the ABI in Ethereum's case) could be stored as, e.g., a JSON structure in the appropriate data source item of the BoM. When the requesting application needs data from the HPC-CS, its runtime application code would unpack the JSON data from the data source item of the BoM and use the address and metadata to interact with the smart contract. The application would write a service request to the HPC-CS's smart contract on the blockchain, laying down an immutable and non-repudiable  record of the request being made for data at a particular time.

It is also possible for the smart contract to take parameters, such as a maximum response time, as previously agreed between the supplier and the customer. In some blockchain implementations, including Ethereum, elements of the smart contract for data provision could be made payable\cite{buterin2014next}, with the data provision service requiring payment before it was delivered. Transactions between requester and data provider which required payment could be processed automatically, with funds being transferred as part of data request transaction written to the blockchain. It is envisaged that data delivered as part of a blockchain-based smart contract request would be sent back to the application asynchronously, and appropriate application logic would be required to receive such data and write its values to the BoL for archival. An experimental system developed alongside this paper has shown that this is feasible. 

To maintain a record of events for ongoing traceability, the structure and elements of the BoM and each BoL would be stored in a database or written to a distributed ledger, such as a blockchain implementation. Where archived runtime data is too large to be feasibly written to a blockchain record it could be written to other decentralised storage systems, for example IPFS\cite{benet2014ipfs}, and a reference to this off-chain storage written in a blockchain transaction. 
Writing the BoM and BoL elements to a public blockchain record would ensure that a permanent data usage history was available, even after projects described by the BoM and BoLs had been disbanded. In a decentralised project, the blockchain record would provide a transaction history with independence of ownership. Participants could agree that the record was correct at the time of writing and the immutable nature of a distributed ledger record would given them assurance that the data recorded could not be changed in the future. To access the details stored on the blockchain in the future, the address or location of the data on the blockchain and the interface required to retrieve that data would need to be archived for future use.

\section{CONCLUSIONS AND FUTURE RESEARCH}
\label{sec:conclusion}

Using a supply chain model to frame analysis of the case studies presented in this paper, has brought benefits in modelling such systems in terms of a BoM into focus. The BoM provides a means to map the overall structure of the elements that make up rich data ecosystems, going beyond the data and considering other contributing factors such as the software and hardware which produces or manages the data, licenses which govern the use and sharing of the data, and policies which contributed to the generation of the data. Instantiating the BoM into a BoL each time the system runs provides a dynamic and traceable view into every invocation of the system, such that the data inputs, data outputs and any artifacts which are used or produced by the system can be archived, readily identified and traced back to their source. Similarly, future users of produced data and artifacts can be identified, which could prove to be very important if errors are later found. Storing metadata capable of identifying smart contracts on a blockchain has enabled immutable recording of the action and timing of requests for data provision, along with the potential for encoding quality of service requirements, and providing automatic payment for services.  

From the current position of our research, we have identified a number of interesting directions that we feel have merit for further research work. As described above, we are interested in extending and deepening the integration of our BoM and BoL models with blockchain technologies such as programmable smart contracts. By associating smart contracts with the data sources and artifacts from the BoM model, we believe that we can facilitate novel dynamic behaviour in the system. Such dynamic behaviours might include runtime selection of the most appropriate data source, along with automatic remuneration and sanctioning, based on dynamic measures of data quality. We are also interested to explore the involvement of human operators in the process, either as staff or crowd-workers, using Decentralised Identifiers\footnote{\url{https://w3c-ccg.github.io/did-spec/}} (DIDs) to associate workers with components of the system and trace their activity and the data and artifacts they are associated with. We believe this may provide a means to reward workers for ongoing use of assets produced by their labours, as advocated by Sriraman, et al\cite{sriraman2017worker}.

We intend to follow the emerging work in the Software BoM community, and to explore how such toolsets might integrate with our own BoM model, particularly where tracking of any software artifacts in our systems is concerned. The progress of the IEEE P7001 Working Group\footnote{\url{http://sites.ieee.org/sagroups-7001/}} on Transparency of Autonomous Systems is also of interest, and we are keen to understand how our work could interplay and how we can use their metrics for transparency to quantify our framework. 

The BoM model offers the potential to map rich data ecosystems and to link components from disparate data systems together to form an overview of the origins and uses of data, along with the accompanying assets and people who contributed to the products. Providing tools that draw upon the data held in the BoM structure to give users easy access to the traceabilty information held within, offers many opportunities to bring visibility and insight to data ecosystems, allowing users to trace the origins and uses of data and to quickly identify any problems. We feel that there is value to be had in exploring this further.

In manufacturing industries, one common use of the BoM for a product is to provide costing data, as it identifies the components which make up an item. In a similar way, the BoM for a collective intelligence system should be able to quantify the cost per instantiation for the system, and bring better visibility onto the usage costs of data-rich systems. We would be interested in exploring ways in which this quantifiable information could be made available to the stakeholders in AI and data ecosystems, such that all participants could better understand both the costs and value creating opportunities in delivering sustainable data ecosystems.

\section*{Acknowledgements}
This research was sponsored by the U.S. Army Research Laboratory and the UK Ministry of Defence under Agreement Number W911NF-16-3-0001. The views and conclusions contained in this document are those of the authors and should not be interpreted as representing the official policies, either expressed or implied, of the U.S. Army Research Laboratory, the U.S. Government, the UK Ministry of Defence or the UK Government. The U.S. and UK Governments are authorized to reproduce and distribute reprints for Government purposes notwithstanding any copyright notation hereon.

\bibliographystyle{spiebib}
\bibliography{datasharing}

\end{document}